\documentclass[preprint,showpacks,preprintnumbers,amsmath,amssymb]{revtex4}

\begin{document}

%\documentclass[]{spie}  %>>> use for US letter paper
%%\documentclass[a4paper]{spie}  %>>> use this instead for A4 paper
%% \addtolength{\voffset}{9mm}   %>>> moves text field down

%  The following command loads a graphics package to include images
%  in the document. It may be necessary to specify a DVI driver option,
%  e.g., [dvips], but that may be inappropriate for some LaTeX
%  installations.
%\usepackage[]{graphicx}

\title{Mutually unbiased phase states, phase uncertainties, and
Gauss sums }

%>>>> The author is responsible for formatting the
%  author list and their institutions.  Use  \skiplinehalf
%  to separate author list from addresses and between each address.
%  The correspondence between each author and his/her address
%  can be indicated with a superscript in italics,
%  which is easily obtained with \supit{}.

\author{Michel Planat}%\supit{a}
\address{{\tt planat@lpmo.edu}\\Institut FEMTO-ST, Departement LPMO,\\ 32
Avenue de
l'Observatoire, 25044 Besan\c{c}on Cedex, France}

\author{Haret Rosu}%\supit{b}
%\skiplinehalf \supit{a}
\address{{\tt hcr@ipicyt.edu.mx}\\
%Div. of Advanced Materials, \\
IPICyT, Apdo Postal 3-74, Tangamanga, San Luis Potos\'{\i},
Mexico }

\bigskip
\bigskip
\bigskip

\author{Dec. 26 2005\\}

\address{}

\bigskip
\bigskip
\bigskip

\address{\Large{{\tt quant-ph/0506128}}}

\vskip 3cm

Europ. Phys. J. D {\bf 36}, 133-139 (Oct. 2005)\hfill PLROJ2.tex

%>>>> Further information about the authors, other than their
%  institution and addresses, should be included as a footnote,
%  which is facilitated by the \authorinfo{} command.

%\authorinfo{Further author information: \\ planat@lpmo.edu, phone: 33 3 81 85 39 57; hcr@ipicyt.edu.mx, phone: +52 4448342000, fax: +52 4448342010}
%%>>>> when using amstex, you need to use @@ instead of @

%%%%%%%%%%%%%%%%%%%%%%%%%%%%%%%%%%%%%%%%%%%%%%%%%%%%%%%%%%%%%
%>>>> uncomment following for page numbers
 \pagestyle{plain}
%>>>> uncomment following to start page numbering at 301
%\setcounter{page}{301}

%  \begin{document}
 % \maketitle

%%%%%%%%%%%%%%%%%%%%%%%%%%%%%%%%%%%%%%%%%%%%%%%%%%%%%%%%%%%%%
\begin{abstract}
Mutually unbiased bases (MUBs), which are such that the inner
product between two vectors in different orthogonal bases is a
constant equal to $1/\sqrt{d}$, with $d$ the dimension of the
finite Hilbert space, are becoming more and more studied for
applications such as quantum tomography and cryptography, and in
relation to entangled states and to the Heisenberg-Weil group of
quantum optics. Complete sets of MUBs of cardinality $d+1$ have
been derived for prime power dimensions $d=p^m$ using the tools of
abstract algebra.
%(Wootters in 1989, Klappenecker in 2003).
Presumably, for non prime dimensions the cardinality is much less.

Here we reinterpret MUBs as quantum phase states, i.e. as
eigenvectors of Hermitean phase operators generalizing those
introduced by Pegg \& Barnett in 1989. We relate MUB states to
additive characters of Galois fields (in odd characteristic p) and
to Galois rings (in characteristic 2). Quantum Fourier transforms
of the components in vectors of the bases define a more general
class of MUBs with multiplicative characters and additive ones
altogether. We investigate the complementary properties of the
above phase operator with respect to the number operator. We also
study the phase probability distribution and variance for general
pure quantum electromagnetic states and find them to be related to
the Gauss sums, which are sums over all elements of the field (or
of the ring) of the product of multiplicative and additive
characters.

Finally, we relate the concepts of mutual unbiasedness and maximal
entanglement. This allows to use well studied algebraic concepts
as efficient tools in the study of entanglement and its
information aspects.

\end{abstract}

%>>>> Include a list of keywords after the abstract

%\keywords{Quantum phase, phase fluctuations, Galois fields,
%mutually unbiased bases}

\maketitle

%%%%%%%%%%%%%%%%%%%%%%%%%%%%%%%%%%%%%%%%%%%%%%%%%%%%%%%%%%%%%
\section{INTRODUCTION}
\label{sect:intro}
%%%%%%%%%%%%%%
%%%%% References %%%%%

In quantum mechanics, orthogonal bases of a Hilbert space
$\mathcal{H}_q$ of finite dimension $q$ are mutually unbiased if
inner products between all possible pairs of vectors of distinct
bases are all equal to $1/\sqrt{q}$.
%They are also said to be maximally non
%commutative in the sense that a measurement over one basis leaves
%one completely uncertain as to the outcome of a measurement
%performed over a basis unbiased to the first.
Eigenvectors of ordinary Pauli spin matrices (i.e. in dimension
$q=2$) provide the best known example.
%With a complete set of $q+1$ mutually unbiased
%measurements one can ascertain the density matrix of en ensemble
%of unknown quantum $q$-states, so that a natural question emerges
%as which mathematics may provide the construction.
It has been shown that in dimension $q=p^m$ which is the power of
a prime $p$, the complete sets of mutually unbiased bases (MUBs)
result from Fourier analysis over a Galois field $F_q$ (in odd
characteristic $p$)\cite{Woott89} or of Galois ring
$R_{4^m}$(in even characteristic $2$)\cite{Klapp03}.
%An exhaustive literature on MUBs can be found
In\cite{Woott04b,Planat04}, one can find an
exhaustive literature on MUBs . Complete sets of MUBs have an
intrinsic geometrical interpretation, and were related to discrete
phase spaces\cite{Woott04b,Paz,Pittinger05},
finite projective planes\cite{Saniga,Saniga2}, convex
polytopes \cite{Bengtson}, and complex projective
$2$-designs\cite{Barnum02,Klap2}. There are hints on the
relation to symmetric informationally complete positive operator
measures
(SIC-POVMs)\cite{Wootters04,Grassl04,Appleby04,Klap05},
%, \cite{Grassl04} , \cite{Appleby04}
and to Latin
squares\cite{Wocjan04}.

There are strong motivations to embark on detailed studies of
MUBs. First, they enter rigorous treatments of Bohr's principle of
complementarity that distinguishes between quantum and classical
systems at the practical level of measurements. This fundamental
quantum principle introduces the idea of complementary pairs of
observables in the sense that precise measurement of one of them
implies that possible outcomes of the other (when measured) are
equally probable. In the nondegenerate case, if an observable $O$
represented by a $q$ times $q$ hermitian matrix is measured in a
quantum system prepared in the eigenbase of its complementary
counterpart $O_c$, then the probability to find the system in one
of the eigenstates of $O$ is just $1/q$ as corresponding to
mutually unbiased inner products. Another domain of applications
where MUBs have been found to play an important role is the field
of secure quantum key exchange (quantum cryptography). In the area
of quantum state tomography, one should use MUBs for a complete
reconstruction of an unknown quantum state \cite{quipro}.

In this paper we approach the MUBs theory from the point of view
of the theory of additive and multiplicative characters in Galois
number field theory. The multiplicative characters
$\psi_k(n)=\exp(\frac{2i\pi n k}{q-1})$, $k=0...q-2$, are well known
since they constitute the basis for the ordinary discrete Fourier
transform. But in order to construct MUBs, the additive characters
introduced below are the ones which are useful. This construction
is implicit in some previous
papers\cite{Woott89,Klapp03,Planat04}, and is now
being fully recognized\cite{Durt04,Klimov04}.

An interesting consequence is the following: the discrete Fourier
transform in $\mathcal{Z}_q$ has been used by Pegg \& Burnett
\cite{Pegg89} as a definition of phase states $|\theta_k \rangle$,
$k=0...q-1$, in $\mathcal{H}_q$. The phase states $|\theta_k
\rangle$ could be considered as eigenvectors of a properly defined
Hermitian phase operator $\Theta _{PB}$. Phase properties and
phase fluctuations attached to particular field states were
extensively described. In particular the classical phase variance
$\pi^2/3$ could be recovered.

We construct here a phase operator $\Theta_{\rm{Gal}}$ having
phase MUBs as eigenvectors. In contrast to the case of $\Theta
_{PB}$, we find that the phase fluctuations of $\Theta_{\rm{Gal}}$
can be expressed in terms of Gauss sums over the finite number
field $F_q$, and could be in principle smaller than those due to
$\Theta _{PB}$. This points to the fact that the phase MUBs may be
of interest for quantum signal processing. Character sums and
Gauss sums which are useful for optimal bases of $m$-qudits ($p$
odd) are also generalized to optimal bases of $m$-qubits ($p=2$).

%It is worthwhile to mention that quadratic Gauss sums were already
%met in the transient and revival dynamics of semi-classical wave
%packets \cite{Berry01} . Finally, related exponential sums:
%Ramanujan sums and Kloosterman sums were found to control the
%phase dynamics of quantum phase-locked states \cite{Planat03} .

%\section{Some character sums over a Galois field}
%\section{Some mathematical concepts}
%\label{sect:Chars}
%%%%%%%%%%%%%%

%Let us consider the field of polynomials $F_p[x]$ defined over the
%field $F_p$
%
%\begin{equation}
%F_p[x]=\{a_0+a_1+\cdots+a_n x^n\},~~a_i\in F_p.
%\end{equation}
%
%For a polynomial $g\in F_p[x]$, the residue class ring
%$F_p[x]/(g)$, where $(g)$ is the ideal class generated by $g$ is a
%field iff g is irreducible over $F_p$ (it cannot be factored over
%$F_p$).

%For example for $q=22$, one can choose the polynomial
%$g(x)=x2+x+1 \in F_2[x]$ which is irreducible over $F_2$.
%Contrary to $\mathcal{Z}_4$ which has zero divisors and is thus
%only a ring, the above construction defines the field with four
%residue classes: $F_4=\{0,1,x,x+1\}$.

%We also mention that more general Gauss sums were studied as
%
%\begin{equation}
%G(\psi,\kappa)=\sum_{x \in F_q}\psi(f(x))\kappa(g(x)),
%\label{Gaussnew}
%\end{equation}
%
%with $f,g \in F_q[x]$ and found to be of the order of magnitude
%$\sqrt{q}$ (\cite{Lidl83} p. 249).

\section{Phase MUBs in odd prime characteristic}
\label{sect:MUBs}
%%%%%%%%%%%%%%

\subsection{Mathematical preliminaries}

The key relation between Galois fields $F_q$ and MUBs  is the
theory of characters. This has not been recognized before and here
we use the standpoint of characters as the most general way of
considering previous results and also as a better criterium for
elaborating on future results.

A Galois field is a finite set structure endowed with two group
operations, the addition \lq\lq$+$" and the multiplication
\lq\lq$\cdot$". The field $F_q$ can be represented as classes of
polynomials obtained by computing modulo an irreducible polynomial
over the ground field $F_p=\mathcal{Z}_p$, the integers modulo
$p$\cite{Lidl83}. A Galois field exists if and only if $q=p^m$. We
also recall that $F_q[x]$ is the standard notation for the set of
polynomials in $x$ with coefficients in $F_q$.

A character $\kappa(g)$ over an abelian group $G$ is a
(continuous) map from $G$ to the field of complex numbers
$\mathcal{C}$ of unit modulus,
%$1$,
i.e. such that $|\kappa(g)|=1$, $g\in G$.

We start with a map from the extended field $F_q$ to the ground
field $F_p$ which is called the trace function
\begin{equation}
tr(x)=x+x^p+\cdots+x^{p^{m-1}}\in F_p,~~ \forall ~x\in F_q.
\label{trace0}
\end{equation}
%

%In addition to its property of mapping an element of $F_q$ into
%$F_p$, the trace function has the properties
%
%\begin{eqnarray}
%&tr(x+y)=tr(x)+ tr(y),~~x,y \in F_q\nonumber \\
%&tr(ax)=atr(x),~~x \in F_q,~a \in F_p,\nonumber \\
%&tr(a)=ma, ~~a\in F_p,\nonumber\\
%&tr(x^q)=tr(x),~~ x\in F_q. \label{trace}
%\end{eqnarray}
%
Using (\ref{trace0}), an additive character over $F_q$ is defined
as
\begin{equation}
\kappa(x)=\omega_p^{tr(x)},~~\omega_p=\exp (\frac{2i\pi}{p}),~~x
\in F_q.
\end{equation}
The main property is that it satisfies
$\kappa(x+y)=\kappa(x)\kappa(y), ~x,y \in F_q$.

On the other hand, the multiplicative characters are of the form
\begin{equation}
\psi_k(n)=\omega_{q-1}^{n k},~~ k=0...q-2, \, n=0...q-2.
\end{equation}

In the present research, the construction of Galois phase MUBs
will be related to character sums with polynomial arguments $f(x)$
also called Weil sums\cite{Klapp03}
\begin{equation}
W_f=\sum_{x \in F_q}\kappa(f(x)). \label{Weilsums}
\end{equation}
In particular, ( theorem 5.38 in \cite{Lidl83}), for a polynomial
$f_d(x) \in F_q[x]$ of degree $d\ge 1$, with $gcd(d,q)=1$, one
gets %$|\sum_{x \in F_q}\kappa(f(x))|
$W_{f_d}\le(d-1)q^{1/2}$.

The quantum fluctuations arising from the phase MUBs will be found
to be related to Gauss sums of the form
\begin{equation}
G(\psi,\kappa)=\sum_{x \in F_q^*}\psi(x)\kappa(x)~, \label{Gauss}
\end{equation}
where $F_q^*=F_q-\{0\}$. Using the notation $\psi_0$ for a trivial
multiplicative character $\psi=1$, and $\kappa_0$ for a trivial
additive character $\kappa=1$ the Gaussian sums (\ref{Gauss})
satisfy $G(\psi_0,\kappa_0)=q-1$; $G(\psi_0,\kappa)=-1$;
$G(\psi,\kappa_0)=0$ and $|G(\psi,\kappa)|=q^{1/2}$ for nontrivial
characters $\kappa$ and $\psi$.

\subsection{Galois quantum phase states}

We now introduce a class of quantum phase states as a \lq\lq
Galois" discrete quantum Fourier transform of the Galois number
kets
\begin{equation}
|\theta^{(y)}\rangle=\frac{1}{\sqrt{q}}\sum_{n \in
F_q}\psi_k(n)\kappa(y n)|n\rangle,~~y \in F_q
 \label{MUB}
\end{equation}
in which the coefficient in the computational base
$\{|0\rangle,|1\rangle,\cdots,|q-1\rangle\}$ represents the
product of an arbitrary multiplicative character $\psi_k(n)$ by an
arbitrary additive character $\kappa(yn)$.

It is easy to show that previous basic results in this area can be
obtained as particular cases of (\ref{MUB}). Indeed:

\noindent \textbf{Pegg \& Barnett (1989)}: For $\kappa=\kappa_0$
and $\psi\equiv \psi_k(n)$, one recovers the ordinary quantum
Fourier transform over $\mathcal{Z}_q$. It has been
shown\cite{Pegg89} that the corresponding states
\begin{equation}
|\theta_k\rangle=\frac{1}{\sqrt{q}}\sum_{n \in
\mathcal{Z}_q}\psi_k(n)|n\rangle, \label{Pegg}
\end{equation}
are eigenstates of the Hermitian phase operator
\begin{equation}
\Theta _{PB}=\sum_{k\in \mathcal{Z}_q}\theta_k|\theta_k\rangle
\langle \theta_k|,
\end{equation}
with eigenvalues $\theta_k=\theta_o+\frac{2\pi k}{q}$, $\theta_0$
an arbitrary initial phase.

\noindent \textbf{Wootters \& Fields (1989)}: We recover the
result of Wootters and Fields in a more general form by employing
the Euclidean division theorem (see theorem 11.19 in
\cite{Lidl98}) for the field $F_q$, which says that given any two
polynomials $y$ and $n$ in $F_q$, there exists a uniquely
determined pair $(a,b)\in F_q \times F_q$, such that $y=an+b$,
$deg(b)<deg(a)$. Using the decomposition of the exponent in
(\ref{MUB}), we obtain
\begin{equation}
|\theta_b^a\rangle=\frac{1}{\sqrt{q}}\sum_{n\in
F_q}\psi_k(n)\kappa(an^2+bn)|n\rangle,~~a,b \in F_q. \label{MUB1}
\end{equation}
(The result of Wootters \& Fields corresponds to the trivial
multiplicative character $\psi_0=1$). Eq.~(\ref{MUB1}) defines a
set of $q$ bases (with index $a$) of $q$ vectors (with index $b$).
Using Weil sums (\ref{Weilsums}) it is easily shown that, for $q$
odd, so that $gcd(2,q)=1$, the bases are orthogonal and mutually
unbiased to each other and to the computational base
\begin{equation}
|\langle \theta_b^a|\theta_d^c\rangle|=|\frac{1}{q}\sum_{n\in F_q}
\omega_p^{tr((c-a)n^2+(d-b)n}|=\left\{\begin{array}{ll}
&\delta_{bd}~~\mbox{if}~c=a~(\mbox{orthogonality}) \\
&\frac{1}{\sqrt{q}} ~~\mbox{if}~c\neq a~(\mbox{unbiasedness}).
\end{array}\right.
\end{equation}

\bigskip

%On the other hand, these phase MUBs are eigenstates of a \lq\lq
%Galois" quantum phase operator
%
%\begin{equation}
%\Theta_{\rm{Gal}}=\sum_{b\in F_q}\theta_b|\theta_b^a\rangle
%\langle \theta_b^a|,~~a,b \in F_q. \label{GalMUB}
%\end{equation}
%
%with eigenvalues $\theta_b=\frac{2\pi b}{q}$. We use this fact to
%perform several calculations of quantum phase expectation values
%and phase variances for these MUBs.

\section{Quantum fluctuations of phase MUBs in odd prime characteristic}
\label{MUBodd}

Following Pegg and Barnett, a good procedure to examine the phase
properties of a quantized electromagnetic field state is by
introducing a phase operator and this was one of the reasons that
led them to introduce their famous Hermitian phase operator
$\Theta _{PB}$. In Section 6 of their seminal paper they showed
``for future reference" how their phase operator could be employed
to achieve this goal. In this section we proceed along the same
lines using the phase form of the Wootters-Field MUBs.

\subsection{The Galois phase operator}

On the other hand, the phase MUBs as given in (\ref{MUB1}) are
eigenstates of a \lq\lq Galois" quantum phase operator
\begin{equation}
\Theta_{\rm{Gal}}=\sum_{b\in F_q}\theta_b|\theta_b^a\rangle
\langle \theta_b^a|,~~a,b \in F_q. \label{GalMUB}
\end{equation}
with eigenvalues $\theta_b=\frac{2\pi b}{q}$. We use this fact to
perform several calculations of quantum phase expectation values
and phase variances for these MUBs.

Using (\ref{MUB1}) in (\ref{GalMUB}) and the properties of the
field theoretical trace %(\ref{trace}),
the Galois quantum phase operator reads
\begin{equation}
\Theta_{\rm{Gal}}=\frac{2\pi}{q^2}\sum_{m,n \in
F_q}\psi_k(n-m)\omega_p^{tr[a(n^2-m^2)]} S(n,m) |n\rangle \langle
m|~,%~\rm{with}~S(n,m)=\sum_{b\in F_q}b\omega_p^{tr[b(n-m)]}.
\label{GalMUB1}
\end{equation}
where $S(n,m)=\sum_{b\in F_q}b\omega_p^{tr[b(n-m)]}$.
In the diagonal matrix elements, we have the partial sums
\begin{equation}
S(n,n)=\frac{q(q-1)}{2}~, \label{Snn}
\end{equation}
so that $\langle n|\Theta_{\rm{Gal}}|n\rangle=\frac{\pi
(q-1)}{q}$. In the non-diagonal matrix elements, the partial sums
can be calculated from
\begin{equation}
\sum_{b\in F_q}b
x^b=x(1+2x+3x^2+\cdots+qx^{q-1})=x\big[\frac{1-x^q}{(1-x)^2}-\frac{qx^q}{1-x}\big]=\frac{xq}{x-1}~,
\end{equation}
where we introduced $x=\omega_p^{tr(n-m)}$ and we made use of the
relation $x^q=1$. Finally, we get
\begin{equation}
S(m,n)=\frac{q}{1-\omega_p^{tr(m-n)}}~. \label{Smn_sums}
\end{equation}

\subsection{The Galois phase-number commutator}  %%% epjd 3.2

Using (\ref{GalMUB1}) and the Galois number operator
\begin{equation}
N=\sum_{l\in F_q}l |l\rangle \langle l|,
\end{equation}
the matrix elements of the phase-number commutator
$[\Theta_{\rm{Gal}},N]$ are calculated as
\begin{equation}
u_{\rm{Gal}}(n,m)=\frac{2\pi}{q^2}(n-m)\psi_k(n-m)\omega_p^{tr[a(n^2-m^2)]}S(n,m).
\end{equation}
The diagonal elements vanish, the corresponding matrix is
anti-Hermitian since
$u_{\rm{Gal}}(n,m)=-u_{\rm{Gal}}^{\dagger}(m,n)$, and the states are
pseudo-classical since $\lim_{q \rightarrow
\infty}u_{\rm{Gal}}(n,m)=0$. These properties are similar to those
of the Pegg \& Barnett commutator.

\subsection{Galois phase properties of a pure quantum electromagnetic state}

For the evaluation of the phase properties of a general pure state
of an electromagnetic field mode in the Galois number field we
proceed similarly to Pegg \& Barnett. Thus, we consider the pure
state of the form
\begin{equation}
|f\rangle=\sum_{n\in
F_q}u_n|n\rangle,~~{\rm with}~u_n=\frac{1}{\sqrt{q}}\exp(i n \beta),
\end{equation}
where $\beta$ is a real parameter, and we sketch the computation
of the phase probability distribution $|<\theta_b|f>|^2$, the
phase expectation value $<\Theta_{\rm{Gal}}>=\sum_{b \in
F_q}\theta_b|<\theta_b|f>|^2$ and the phase variance
$<\Delta\Theta_{\rm{Gal}}^2>=\sum_{b \in
F_q}(\theta_b-<\Theta_{\rm{Gal}}>)^2|<\theta_b|f>|^2$,
respectively
%
%\begin{eqnarray}
%&|<\theta_b|f>|^2, \nonumber \\
%& <\Theta_{\rm{Gal}}>=\sum_{b \in F_q}\theta_b|<\theta_b|f>|^2,
%\nonumber \\
%&<\Delta\Theta_{\rm{Gal}}^2>=\sum_{b \in
%F_q}(\theta_b-<\Theta_{\rm{Gal}}>)2|<\theta_b|f>|^2,
%\end{eqnarray}
%
(the upper index $a$ for the base is implicit and we discard it
for simplicity).

%\subsection{Phase expectation value}

The two factors in the expression for the probability distribution
\begin{equation} %%%%  Eq. (19)
 \frac{1}{q^2} [\sum_{n \in F_q} \psi_k(-n)
\exp(i n \beta)\kappa (-an^2-bn)][\sum_{m \in F_q} \psi_k(m)
\exp(-i m \beta)\kappa (am^2+bm)],
\end{equation}
have absolute values bounded by the absolute value of generalized
Gauss sums
%\begin{equation}
$G(\psi,\kappa)=\sum_{x \in F_q}\psi(g(x))\kappa(f(x))$,
%\label{Gaussnew}
%\end{equation}
%
with $f,g \in F_q[x]$. Weil \cite{W48} showed that for $f(x)$ of
degree $d$ with $gcd(d,q)=1$ as in (\ref{Weilsums}), under the
constraint that for the multiplicative character $\psi$ of order
$s$, the polynomial $g(x)$ should not be a $s$th power in $F_q[x]$
and with $\nu$ distinct roots in the algebraic closure of $F_q$,
the order of magnitude of the sums is $(d+\nu -1)\sqrt{q}$.
%and found to be of the order of magnitude
%$\sqrt{q}$ (\cite{Lidl83} p. 249). Thus,
For a trivial multiplicative character $\psi _0$, and $\beta =0$,
the overall bound is $|<\theta_b|f>|^2\le \frac{1}{q}$ and it
follows that the absolute value of the Galois phase expectation
value is bounded from above as expected for a common phase operator
%..........................
\begin{equation}
|<\Theta_{\rm{Gal}}>|\le \frac{2\pi}{q^2}\sum_{b \in F_q}b\le \pi.
\end{equation}
%...........................
The exact formula for the phase expectation value reads
\begin{equation}
<\Theta_{\rm{Gal}}>=\frac{2 \pi}{q^3}\sum_{m,n \in
F_q}e^{\beta}(m,n)
%\psi_k(m-n)\exp[i(n-m)\beta]\omega_p^{tr[a(m2-n2]}
S(m,n), \label{expect}
\end{equation}
where $e^{\beta}(m,n)= \psi_k(m-n)\exp[i(n-m)\beta]\chi
[a(m^2-n^2)]$ and the sums $S(m,n)$ were defined in (\ref{Snn})
and (\ref{Smn_sums}). The set of all the $q$ diagonal terms $m=n$
in $<\Theta_{\rm{Gal}}>$ contributes an order of magnitude
$\frac{2\pi}{q^3}q S(n,n)\simeq \pi$. The contribution from
off-diagonal terms in (\ref{expect}) are not easy to evaluate
analytically; we were able to show that for them one has
$|S(m,n)|=\frac{q}{2}|\sin [\frac{\pi}{p}tr(n-m)]|^{-1}$.

%and possible cancellation of phase oscillations could be
%considered from numerical plots, since the sums in (\ref{expect})
%are not easy to evaluate analytically.

%\subsection{Phase variance}

The phase variance can be written as
\begin{equation}
<\Delta\Theta_{\rm{Gal}}^2>=\sum_{b \in F_q}(\theta_b^2-2 \theta_b
<\Theta_{\rm{Gal}}>)|<\theta_b|f>|^2.
 \label{variance}
\end{equation}
The term $<\Theta_{\rm{Gal}}>^2\sum_{b\in F_q}|<\theta_b|f>|^2$
does not contribute since it is proportional to the Weil sum
$\sum_{b \in F_q}\omega_p^{tr(b(n-m)}=0$. As a result a
cancellation of the quantum phase fluctuations may occur in
(\ref{variance}) from the two extra terms of opposite sign. But
the calculation are again not easy to perform analytically. For
the first term one gets
%
%\begin{equation}
$ 2( 2\pi/q^2)^2 \sum_{m,n \in F_q}e^{\beta}(m,n) |S(m,n)|^2~.
%,~~\rm{with}~T(n,m)=\sum_{b\in F_q}b2\omega_p^{tr[b(n-m)]}.
$
The second term in (\ref{variance}) is
%
%\begin{equation}
$ -2\sum_{b \in F_q}\theta_b
<\Theta_{\rm{Gal}}>|<\theta_b|f>|^2=-2<\Theta_{\rm{Gal}}>^2
%\label{SecondTerm}
%\end{equation}
$.
Partial cancellation occurs in the diagonal terms of
(\ref{variance}) leading to the contribution $\approx
-\frac{2\pi^2}{3}$
%
%\begin{equation}
%\frac{4\pi2}{q4}q T(n,n)-\frac{8 \pi2}{q4}S(n,n)2\simeq
%\frac{4\pi2}{3}-2\pi2=-\frac{2\pi2}{3},
%\end{equation}
%
which is still twice (in absolute value) the amount of phase
fluctuations in the classical regime. A closed form for the
estimate of the non-diagonal terms is still an open problem.

\section{Phase MUBs for ${\rm m}$-qubits}
\label{sect:MUB2s}

\subsection{Mathematical preliminaries}

 The Weil sums (\ref{Weilsums}) which have been proved useful
in the construction of MUBs in odd characteristic $p$ (and odd
dimension $q=p^m$), are not useful in characteristic $p=2$, since
in this case the degree $2$ of the polynomial $f_d(x)$ is such
that $gcd(2,q)$=2.

An elegant method for constructing complete sets of MUBs of
$m$-qubits was found by Klappenecker and R\"otteler\cite{Klapp03}.
It makes use of objects belonging to the context of quaternary
codes \cite{Wan97}, the so-called Galois rings $R_{4^m}$; we refer
the interested reader to their paper for more mathematical
details. We present a brief sketch in the following.

Any element $y \in R_{4^m}$ can be uniquely determined in the form
$y=a + 2 b$, where $a$ and $b$ belong to the so-called
Teichm\"{u}ller set $\mathcal{T}_m =
(0,1,\xi,\cdots,\xi^{2^m-2})$, where $\xi$ is a nonzero element of
the ring which is a root of the so-called basic primitive
polynomial $h(x)$ \cite{Klapp03}. Moreover, one finds that
$a=y^{2^m}$. We can also define the trace to the base ring
$\mathcal{Z}_4$ by the map
\begin{equation}
\tilde{tr}(y)=\sum_{k=0}^{m-1}\sigma^k(y), \label{trace2}
\end{equation}
where the summation runs over $R_{4^m}$ and the Frobenius
automorphism $\sigma$ reads
\begin{equation}
\sigma(a+2 b)=a^2+ 2 b^2.
\end{equation}

In the Galois ring of characteristic $4$ the additive characters
are
\begin{equation}
\tilde{\kappa}(x)=\omega_4^{\tilde{\rm{tr}}(x)}=i^{\tilde{\rm{tr}}(x)}.
\end{equation}
%
%\subsection{Exponential sums over $R_{4^m}$}

The Weil sums (\ref{Weilsums}) are replaced by the exponential
sums \cite{Klapp03}
\begin{equation}
\Gamma(y)=\sum_{u \in \mathcal{T}_m}\tilde{\kappa}(y u),~~y \in
R_{4^m} \label{newWeilsums}
\end{equation}
which satisfy
\begin{equation}
|\Gamma(y)|=\left\{\begin{array}{ll}
&0~~\mbox{if}~y \in 2 \mathcal{T}_m,~y \ne 0 \\
&2^m ~~\mbox{if}~y=0\\
&\sqrt{2^m}~~\mbox{otherwise}.
\end{array}\right.
\end{equation}
Gauss sums for Galois rings were constructed \cite{Oh01}
\begin{equation}
G_y(\tilde{\psi},\tilde{\kappa})=\sum_{x \in R_{4^m}}
\tilde{\psi}(x)\tilde{\kappa}(yx), ~~y \in
R_{4^m},\label{GaussGal}
\end{equation}
where the multiplicative character $\bar{\psi}(x)$ can be made
explicit \cite{Oh01}.

Using the notation $\bar{\psi_0}$ for a trivial multiplicative
character and $\tilde{\kappa_0}$ for a trivial additive character,
the Gaussian sums (\ref{GaussGal}) satisfy
$G_y(\tilde{\psi_0},\tilde{\kappa_0})=4^m$;
$G_y(\tilde{\psi},\tilde{\kappa_0})=0$ and
$|G_y(\tilde{\psi},\tilde{\kappa})|\le 2^m$.

\subsection{Phase states for $m$-qubits}

The quantum phase states for $m$-qubits can be found as the \lq\lq
Galois ring" Fourier transform

\begin{equation}
|\theta^{(y)}\rangle=\frac{1}{\sqrt{2^m}}\sum_{n \in
\mathcal{T}_m}\tilde{\psi}_k(n)\tilde{\kappa}(y n)|n\rangle,~~y
\in R_{4^m}.
 \label{MUBpair}
\end{equation}
%

%\subsection{A. Klappenecker \& M. R\"{o}tteler 03}

Using the Teichm\"{u}ller decomposition in the character function
$\tilde{\kappa}$ one obtains
\begin{equation}
|\theta_b^a\rangle=\frac{1}{\sqrt{2^m}}\sum_{n\in
\mathcal{T}_m}\tilde{\psi}_k(n)\tilde{\kappa}[(a+2b)n]|n\rangle,~~a,b
\in \mathcal{T}_m. \label{MUB1pair}
\end{equation}
This defines a set of $2^m$ bases (with index $a$) of $2^m$
vectors (with index $b$). Using the exponential sums
(\ref{newWeilsums}), it is easy to show that the bases are
orthogonal and mutually unbiased to each other and to the
computational base. The case $\bar{\psi}\equiv \bar{\psi_0}$ was
obtained before \cite{Klapp03}.

\subsection{Phase MUBs for m-qubits: $m=1$, $2$ and $3$}

For the special case of qubits, one uses $\tilde{\rm{tr}}(x)=x$ in
(\ref{MUB1pair}) so that the three pairs of MUBs are given as
\begin{equation}
[|0\rangle,|1\rangle];~~\frac{1}{\sqrt{2}}[|0\rangle+|1\rangle,|0\rangle-|1\rangle];~~\frac{1}{\sqrt{2}}[|0\rangle+
i |1\rangle,|0\rangle- i |1\rangle]. \nonumber
\end{equation}

For $2$-qubits one gets a complete set of $5$ bases as follows
\begin{eqnarray}
&(|0\rangle,|1\rangle,|2\rangle,|3\rangle); \nonumber\\
&\frac{1}{2}[|0\rangle+|1\rangle+|2\rangle+|3\rangle,|0\rangle+|1\rangle-|2\rangle-|3\rangle,
 |0\rangle-|1\rangle-|2\rangle+|3\rangle,|0\rangle-|1\rangle+|2\rangle-|3\rangle]\nonumber\\
&\frac{1}{2}[|0\rangle-|1\rangle-i|2\rangle-i|3\rangle,|0\rangle-|1\rangle+i|2\rangle+i|3\rangle,
 |0\rangle+|1\rangle+i|2\rangle-i|3\rangle,|0\rangle+|1\rangle-i|2\rangle+i|3\rangle]\nonumber\\
 &\frac{1}{2}[|0\rangle-i|1\rangle-i|2\rangle-|3\rangle,|0\rangle-i|1\rangle+i|2\rangle+|3\rangle,
 |0\rangle+i|1\rangle+i|2\rangle-|3\rangle,|0\rangle+i|1\rangle-i|2\rangle+|3\rangle]\nonumber\\
 &\frac{1}{2}[|0\rangle-i|1\rangle-|2\rangle-i|3\rangle,|0\rangle-i|1\rangle+|2\rangle+i|3\rangle,
 |0\rangle+i|1\rangle+|2\rangle-i|3\rangle,|0\rangle+i|1\rangle-|2\rangle+i|3\rangle],\nonumber\\
 \label{quartrits}
\end{eqnarray}

and for $3$-qubits a complete set of $9$ bases
\begin{eqnarray}
&(|0\rangle,|1\rangle,|2\rangle,|3\rangle,|4\rangle,|5\rangle,|6\rangle,|7\rangle); \nonumber\\
&\frac{1}{4}[|0\rangle+|1\rangle+|2\rangle+|3\rangle+|4\rangle+|5\rangle+|6\rangle+|7\rangle,
|0\rangle+|1\rangle-|2\rangle+|3\rangle-|4\rangle-|5\rangle-|6\rangle+|7\rangle,\nonumber\\
&|0\rangle-|1\rangle+|2\rangle-|3\rangle-|4\rangle-|5\rangle+|6\rangle-|7\rangle,
|0\rangle+|1\rangle-|2\rangle-|3\rangle-|4\rangle+|5\rangle+|6\rangle-|7\rangle,\nonumber\\
&|0\rangle-|1\rangle-|2\rangle-|3\rangle+|4\rangle+|5\rangle-|6\rangle+|7\rangle,
|0\rangle-|1\rangle-|2\rangle+|3\rangle+|4\rangle-|5\rangle+|6\rangle-|7\rangle,\nonumber\\
&|0\rangle-|1\rangle+|2\rangle+|3\rangle-|4\rangle+|5\rangle-|6\rangle-|7\rangle,
|0\rangle+|1\rangle+|2\rangle-|3\rangle+|4\rangle-|5\rangle-|6\rangle-|7\rangle],\nonumber\\
&\cdots
\end{eqnarray}
where only the first two bases have been written down for brevity
reasons.

%\subsection{Quantum phase fluctuations for $m$-qubits}
Quantum phase states of $m$-qubits (\ref{MUB1pair}) are
eigenstates of a \lq\lq Galois ring" quantum phase operator as in
(\ref{GalMUB}), and calculations of the same type as to those
performed in Sect. (\ref{MUBodd}) can be done, since the
$\tilde{\rm{tr}}$ operator (\ref{trace2}) fulfills rules similar
to the $\rm{tr}$ operator (\ref{trace0}). By analogy to the case
of qudits in dimension $p^m$, $p$ an odd prime, phase properties
for sets of $m$-qubits heavily rely on the Gauss sums
(\ref{GaussGal}). The calculations are tedious once again but can
in principle be achieved in specific cases.

\section{Mutual unbiasedness and maximal entanglement}

%It has been shown in this paper that there is a founding link
%between irreducible polynomials over a ground field $F_p$ and
%complete sets of mutually unbiased bases arising from Fourier
%transform over a lifted field $F_q$, $q=p^m$, $p$ a prime number.
%On the other hand, the physical concept of entanglement over the
%Hilbert space $\mathcal{H}_q$ evokes irreducibility.

Roughly speaking, entangled states in $\mathcal{H}_q$ cannot be
factored into tensorial products of states in Hilbert spaces of
lower dimension. We show now that there is an intrinsic relation
between MUBs and maximal entanglement (see below).

We start with the familiar Bell states
\begin{eqnarray}
&(|\mathcal{B}_{0,0}\rangle,|\mathcal{B}_{0,1}\rangle)=\frac{1}{\sqrt{2}}(|00\rangle+|11\rangle,|00\rangle-|11\rangle),
\quad %\nonumber\\
&(|\mathcal{B}_{1,0}\rangle,|\mathcal{B}_{1,1}\rangle)=\frac{1}{\sqrt{2}}(|01\rangle+|10\rangle,|01\rangle-|10\rangle),
\nonumber
\end{eqnarray}
where the compact notation $|00\rangle=|0\rangle\odot|0\rangle$,
$|01\rangle=|0\rangle\odot|1\rangle$,\dots, is employed for the
tensorial products.

These states are both orthonormal and maximally entangled, i.e.,
such that $trace_2|\mathcal{B}_{h,k}\rangle
\langle\mathcal{B}_{h,k}| =\frac{1}{2}I_2$, where $trace_2$ means
the partial trace over the second qubit \cite{Nielsen00}.

One can define more general Bell states using the multiplicative
Fourier transform (\ref{Pegg}) applied to the tensorial products
of two qudits \cite{Cerf01}\cite{Durt04},
\begin{equation}
|\mathcal{B}_{h,k}\rangle=\frac{1}{\sqrt{q}}\sum_{n=0}^{q-1}\omega_q^{k
n}|n,n+h\rangle, \label{FourierEntang}
\end{equation}
These states are both orthonormal, $\langle
\mathcal{B}_{h,k}|\mathcal{B}_{h',k'} \rangle
=\delta_{hh'}\delta_{kk'}$, and maximally entangled,
$trace_2|\mathcal{B}_{h,k}\rangle \langle\mathcal{B}_{h,k}|
=\frac{1}{q}I_q$.

We define here an even more general class of maximally entangled
states using the Fourier transform (\ref{MUB1}) over $F_q$ as
follows
\begin{equation}
|\mathcal{B}_{h,b}^a\rangle=\frac{1}{\sqrt{q}}\sum_{n=0}^{q-1}\omega_p^{tr[(a
 n + b)n ]}|n,n+ h\rangle ~.
\label{entangledGalois}
\end{equation}
The $h$ we use here has nothing to do with the polynomial $h(x)$
of Section (\ref{sect:MUBs}). A list of the generalized Bell
states of qutrits for the base $a=0$ can be found in
\cite{Fujii01} which is a work that relies on a coherent state
formulation of entanglement. In general, for $q$ a power of a
prime, starting from (\ref{entangledGalois}) one obtains $q^2$
bases of $q$ maximally entangled states. Each set of the $q$ bases
(with $h$ fixed) has the property of mutual unbiasedness.

Similarly, for sets of maximally entangled m-qubits one uses the
Fourier transform over Galois rings (\ref{MUB1pair}) so that
\begin{equation}
|\mathcal{B}_{h,b}^a\rangle=\frac{1}{\sqrt{2^m}}\sum_{n=0}^{2^m-1}i^{tr[(a
+  2 b) n]}|n,n+ h\rangle \label{twoevendits}.
\end{equation}
For qubits ($m=1$) one gets the following bases of maximally
entangled states (in matrix form, up to the proportionality
factor)
\begin{equation}
 \left [\begin{array}{cc} (|00\rangle+|11\rangle,|00\rangle-|11\rangle) &(|01\rangle+|10\rangle,|01\rangle-|10\rangle)
 \\(|00\rangle+i|11\rangle,|00\rangle-i|11\rangle) &(|01\rangle+i|10\rangle,|01\rangle)-i|10\rangle)\\
\end{array}\right].
\end{equation}
Two bases in one column are mutually unbiased, while vectors in
two bases on the same line are orthogonal to each other.

For two-particle sets of quartits, using Eqs.~(\ref{quartrits})
and (\ref{twoevendits}), one gets $4$ sets of
$(|\mathcal{B}_{h,b}^a\rangle$, $ h=0,...,3)$, see them below,
each entailing $4$ MUBs $(a=0,...,3)$:
%\medskip
%{\it h=0 set}
\begin{eqnarray}
&\{(|00\rangle+|11\rangle+|22\rangle+|33\rangle \quad |++--\rangle \quad %\nonumber
|+--+\rangle \quad |+-+-\rangle);\nonumber\\
&(|00\rangle-|11\rangle-i|22\rangle-i|33\rangle \quad |+-(+i)(+i)\rangle \quad%\nonumber\\
|++(+i)(-i)\rangle \quad
|++(-i)(+i)\rangle) ; \cdots\nonumber
\} \nonumber
\end{eqnarray}
%................................
\begin{eqnarray}
&\{(|01\rangle+|12\rangle+|23\rangle+|30\rangle \quad |++--\rangle \quad %\nonumber\\
|+--+\rangle \quad |+-+-\rangle); \nonumber\\
&(|01\rangle-|12\rangle-i|23\rangle-i|30\rangle \quad |+-(+i)(+i)\rangle \quad %\nonumber\\
|++(+i)(-i)\rangle \quad
|++(-i)(+i)\rangle) ; \cdots\nonumber
\} \nonumber%\nonumber\\
%\cdots \} \nonumber\}
\end{eqnarray}
$$
\{(|02\rangle+|13\rangle+|20\rangle+|31\rangle \quad |++--\rangle \quad %\nonumber\\
|+--+\rangle \quad |+-+-\rangle);\cdots %\nonumber\\
%&\cdots\nonumber
\}
%\end{eqnarray}
$$
%\begin{eqnarray}
%\begin{equation}
%..........................
%\medskip
%{\it h=3 set}
$$
%&
\{(|03\rangle+|10\rangle+|21\rangle+|32\rangle\quad
|++--\rangle\quad
|+--+\rangle\quad
|+-+-\rangle); \cdots \}
%\nonumber\\
%&\cdots
%\},
%\end{eqnarray}
%\end{equation}
$$
where, for the sake of brevity, we omitted the normalization
factor ($1/2$) and the bases in the sets have been labeled by their coefficients unless for the first base.
Thus, in the first set $|00\rangle+|11\rangle+|22\rangle+|33\rangle\equiv |++++\rangle$. Within each set, the four bases are mutually
unbiased, as in (\ref{quartrits}), while the vectors of the bases
from different sets are orthogonal.

As a conclusion, the two related concepts of mutual unbiasedness
and maximal entanglement derive from the study of lifts of the
base field $\mathcal{Z}_p$ to Galois fields of prime
characteristic $p>2$ (in odd dimension), or of lifts of the base
ring $\mathcal{Z}_4$ to Galois rings of characteristic $4$ (in
even dimension). One wonders if lifts to more general algebraic
structures would play a role in the study of non maximal
entanglement. We have first in mind the nearfields that are used
for deriving efficient classical codes and which have a strong
underlying geometry\cite{gpiltz}.

\section{Conclusion}

In this research, we approached the MUBs fundamental topic from the
point of view of the additive and multiplicative characters over
finite fields in number theory. We consider that this framework is
the most general including previous results in the literature as
particular cases. Since MUBs are essentially generalized discrete
Fourier transforms over finite number field kets, we formulated a
quantum phase interpretation and illustrated several calculations of
the phase properties of pure quantum states of the electromagnetic
field in this finite number field mathematical context. Various
types of Gauss sums get involved in this type of calculations of the
MUBs phase properties of a pure quantum state and the generalization
to the mixed states, although straightforward through the usage of
the density matrix formalism, could lead to even more complicated
calculations involving such sums. We hope to evaluate them in future
works. We also mentioned in the last section a possible application
to phase MUBs states of Bell type. This could lead to finite number
field measures of the degree of entanglement.

\bigskip
\bigskip

\noindent {\bf Note}: The authors acknowledge Dr. Igor Shparlinski
for suggesting important corrections on this paper. On June 20/2005
he sent us an e-message pointing to some mathematical
inconsistencies in subsection III.C (i.e., subsection 3.3 in the
epjd version). Parameter $\beta$ therein should be itself an element
in $F_q$ to perform the calculations.
%Dr. Planat proposed that Eq.~(19) be
%eliminated but nevertheless it appeared in the published version.

%\bibliography{reportnew}   %>>>> bibliography data in report.bib
%\bibliographystyle{spiebibnew}   %>>>> makes bibtex use spiebib.bst

\end{document}